\documentstyle[figspecial,psfig,conf_iap,art10pt]{article}
\begin{document}
\def\ga{\lower 2pt \hbox{$\, \buildrel {\scriptstyle >}\over{\scriptstyle \sim}\,$}}
\def\la{\lower 2pt \hbox{$\, \buildrel {\scriptstyle <}\over{\scriptstyle \sim}\,$}}
\heading{Ly $\alpha$ Cloud Size Evolution and Shape from QSO Groups}
\par\medskip\noindent

\author{Arlin P.S.~Crotts$^1$}
\address{$^1$Dept.~of Astronomy, Columbia University, 550 West 120 St., New York, NY, U.S.A.}

\begin{abstract}{\baselineskip 0.4cm 
We present results from Keck HIRES, KPNO 4m/RC Spec and $HST$ FOS spectra of a
low-redshift, close QSO pair and a close, higher-redshift QSO triplet in order
to explore the size evolution, shape and clustering of Ly $\alpha$ absorbers.
The results we highlight here indicate that there is no strong evidence for
evolution of Ly $\alpha$ cloud size with redshift, that strong ($W_o >
0.4$\AA) Ly $\alpha$ forest absorbers are probably not filaments, but probably
more sheet-like.
For C~IV absorbers a large component of clustering
on 200~km~s$^{-1} < \Delta v < 600$~km~s$^{-1}$ scales along single
sightlines is due to internal velocities of single absorbers, not clustering
between absorbers.
}
\end{abstract}

\section{Introduction}

Over the past several years we have embarked on a concentrated effort to use
multiple QSO sightlines to reveal new insights into QSO absorption systems,
both C IV and Ly~$\alpha$.
These include observations of the 9.5-arcsec wide pair Q1343+2640A/B ($z =
2.05$ for both QSOs) obtained at the MMT, plus observations of the wider pair
at 1517+239 (at $z \approx 1.9$) and QSO triplet 1623+269 (at $z \approx 2.5$)
with the KPNO 4-meter RC Spec, and of 1623+269 with Keck HIRES.
In this paper we will summarize a few results (indicated by an asterisk) of the
1623+269 and 1517+239 studies, found in our most recent papers [3,4], but
would also like to mention some of the results from other works in the series:

\medskip

\noindent From the Q1343+2640A/B pair:

$-$ Ly~$\alpha$ forest clouds have a radius $R \approx 100 ~h^{-1}$kpc (for $W_o \ga 0.4$\AA) [1,2].

$-$ Ly~$\alpha$ clouds are not minihalos, pressure-confined or freely-expanding [1,2].

$-$ Ly~$\alpha$ cloud number and mass are like faint blue galaxies' (FBGs) [1,2].

$-$ C IV absorbers are on the order of $40~h^{-1}$kpc in
radius [1,2].

$-$ Baryonic mass in forest $\ga$ 
baryons in Ly limit or damped Ly $\alpha$ clouds \cite{FC}.

$-$ Ly~$\alpha$ forest contains a large fraction of baryons in Universe at $z
\approx 2$~~\cite{FC}.

$-$ Re-estimate: $W_o \ge 0.4$\AA\ Ly~$\alpha$ clouds have
$R \approx 150 ~h^{-1}$kpc \cite{FDCB}.

$-$ Ly~$\alpha$ clouds can be modelled as collapsing to form FBGs
\cite{FDCB}.

$-$ We study differences in $W_o$, velocity centroid across sightlines \cite{FDCB}.

$-$ Ly~$\alpha$ clouds cannot be unclustered,
uniform-sized spheres \cite{FDCB}.

\noindent From Q1623+269 triplet and Q1517+239 pair \cite{CF}:

$-$ There is no significant evidence for Ly~$\alpha$ cloud size
evolution$^*$.

$-$ Large Ly~$\alpha$ clouds are not consistent with filaments$^*$.

$-$ Ly~$\alpha$ cloud kinematics and spatial uniformity suggest gas sheets$^*$.

$-$ There are weak signs of large voids in the Ly~$\alpha$
forest.

$-$ The background/foreground QSO proximity effect is tested.

\noindent From Keck HIRES data on Q1623+269 triplet \cite{CBT}:

$-$ C IV lines clustering across sightlines is weaker than in single
sightlines$^*$.

\medskip

First, we make a few comments the results we will {\it not} discuss in detail.
The model of Ly~$\alpha$ clouds as progenitors of FBGs
is supported by the size estimate for Ly~$\alpha$ clouds from
Q1343+2640, which allows one to also calculate the comoving spatial number
density and neutral hydrogen mass (making assumptions about cloud shape),
revealing several similarities between the populations: nearly
identical (comoving) spatial number densities (about 0.3~Mpc$^{-3}$, with
$H_o = 100$~km~s$^{-1}$~Mpc$^{-1}$), both very high
in comparison to other extragalactic populations.
They also have similar masses and clustering strengths.
The size and mass of Ly~$\alpha$ clouds give an estimate of
their collapse times (also given their quiescent internal velocities), which at
$z \approx 2$ correspond to collapse at $z \approx 1$, consistent
with constraints on the formation epoch of FBGs.

We note \cite{FC} that the Ly~$\alpha$ cloud size determination
(along with their line-of-sight number density) allows the contribution of
Ly~$\alpha$ clouds to the critical density, $\Omega_{Ly\alpha}$, to be
estimated directly.
It is larger or roughly equal to the baryon mass in Ly limit or damped
Ly~$\alpha$, which altogether compose most of the baryons produced in the Big
Bang.
Rauch \& Haehnelt \cite{RH} find a similar result.
These determinations depend on much simpler assumptions and are independent of
estimates of $\Omega_{baryon}$ obtained by comparing absorption structure along
single sightlines to numerical models \cite{Retal}.
We now consider several recent results [3,4]:

\section{Results and Interpretation:}
\subsection{Evolution of Ly $\alpha$ Cloud Size?}

The QSO pair (1517+2357 at $z = 1.834$
and 1517+2356 at $z = 1.903$, with 432~$h^{-1}$~kpc transverse proper
separation) and the QSO triplet KP 76/KP 77/KP 78 (1623+2651A at $z = 2.467$,
1623+2653 at $z = 2.526$, and 1623+2651B at $z = 2.605$, respectively, with
proper transverse separations of $0.5-0.7~h^{-1}$Mpc) were observed.
Data on all five QSOs using the RC Spectrograph on the KPNO 4-meter were
obtained at 1.4-1.7\AA\ FWHM resolution by us and Elowitz et al.~\cite{EGI}
(for 1517+2356/7), and for 1517+2356/7 on $HST$
using the FOS G190H and G270H setups as part of program GO 5320 of Foltz et al.
We use sightline cross-correlations of these data to estimate cloud size.
For velocity differences $\Delta v < 200$~km~s$^{-1}$ between lines in adjacent
sightlines, one sees a large surplus for the QSO triplet (21 seen
versus 7.3 expected, random at the $10^{-5}$ probability level)
on $0.5-1.0~h^{-1}~$Mpc scales.
This surplus persists over other published pairs [5,12], and
we refer to such a $\Delta v < 200$~km~s$^{-1}$ coincidence as a ``hit.''

Do the sizes of Ly $\alpha$ clouds change over cosmic time?
1517+2356/7 is valuable in this connection,
being intermediate in redshift between 0107-0234/5, at $z
\approx 0.95$, and other pairs at $z \approx 2$ or higher.
Since it was observed with $HST$, its useful Ly $\alpha$ range extends to
lower redshifts compared to the $z \ga 2$ sample.


\begin{figure}
\plotfiddle{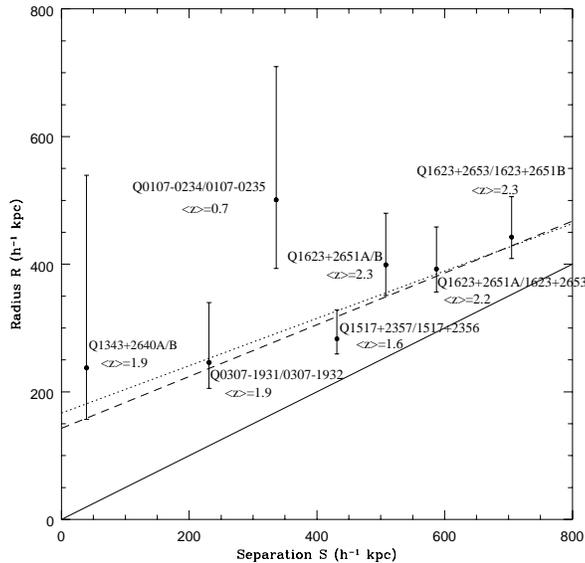}{1.4in}{0}{40}{40}{-120}{-80}
\caption{The inferred radius of Ly~$\alpha$ clouds versus pair separation.}
\end{figure}

Figure 1 shows the inferred size of $W_o>0.4$\AA\ Ly~$\alpha$ clouds (assuming
that they are uniform-sized, unclustered spheres) for 1517+2356/7 in
relation to 0107-0234/5 and higher-redshift pairs.
(Shown are the median size calculated from the Bayesian probability
distribution, assuming uniform priors in cloud radius, and the confidence
intervals corresponding to $\pm 1 \sigma$ error bars.)~
These calculations are made according to our previous procedure \cite{FDCB},
and include our new data on the 1623+269 triplet.
As discovered previously \cite{FDCB}, there is a significant trend of median
estimated $R$ with $S$, contrary to our assumed model.
The slope in a linear fit of $R$ versus $S$ is $0.37 \pm 0.18$ for all QSO
pairs (dashed line).
If, noting that 0107-0234/5 appears to be discrepant, one leaves it
out, one finds that the trend of $R$ with $S$ is almost unchanged and equally
significant, with slope of $0.41 \pm 0.18$ (dotted line).
The other lower redshift QSO pair, 1517+2356/7, falls {\it below} the
trend set by higher redshift QSOs.
With the $R(S)$ dependence removed, one finds the 0107-0234/5 point
sitting $1.2 \sigma$ above the minimum $\chi^2$ linear fit of $R$ versus $z$,
which shows no significant trend of size increase with $z$ (best fit
$\partial R/\partial z = -13$ kpc per unit $z$, with an error of 81~kpc per
unit $z$).
There is so strong evidence for size evolution.

\subsection{Sheet-like or Filamentary Ly $\alpha$ Clouds?}

It is also possible to use ``hit'' statistics to test directly the
non-spherical models.
Numerical models of intergalactic objects in the early Universe 
tend to find elongated structures on the scale of several hundred
kpc as those with properties most similar to Ly $\alpha$ clouds.
The triplet is ideal for determining whether the hit statistics deviate from an
$S$-independent $R$ due primarily to elongated clouds; single, long, thin
filaments are incapable of intercepting all three sightlines.
Such an effect should be expressible as the probability of clouds of a given
shape and size hitting two or all three sightlines whenever they hit one.
This is accomplished by, first, measuring in the actual spectra's linelists the
probabilities $P_{ab}$, $P_{ac}$, $P_{bc}$ and $P_{abc}$, (defined as $P_{ab}$
being the probability of a line in A resulting in a hit in B, or vice versa,
and likewise for the other probabilities) and, secondly, simulating the same
probabilities by a numerical simulation of cylindrical rods of various aspect
ratio $a$ and cross-sectional radius $R$ values.
The probabilities are computed for each rod shape and size
and agree better for small $a$.
The statistical significance of the difference between large and small
$a$ is not great, decreasing from a maximum $1.2 \sigma$ for large $a$ to
0.6-0.7$\sigma$ for $a < 2$.
Nevertheless, all probabilities have best agreement for $1 < a < 3$ and
$198~h^{-1}$kpc$~ < R < 510~h^{-1}$kpc (larger $R$ at smaller $a$), where
$\sigma < 0.8$.
The high probability of three-way hits argues for sheets rather than filaments.

We summarize another significant result:
when we now consider how the difference in linestrength $|W_A - W_B|$
compares to the maximum linestrength $W_A$ (or $W_B$), we find that the most
uniform lines (smallest $|W_A - W_B|$ at given $max(W_A,W_B)$ occur almost
exclusively in the small fraction of lines {\it which span all three sighlines
in the QSO triplet}.
Clouds which have an extent of at least $0.5~h^{-1}$~Mpc in two dimensions are
also the most uniform in neutral hydrogen strength, suggesting sheets.

\subsection {C IV Internal Structure for $\Delta v=200-600$~km s$^{-1}$}

Clustering of C~IV lines in the QSO triplet do not show the
strong signal in cross-correlation between sightlines seen in auto-correlations
along single sightlines.
For instance, for lines with $W_o(1548$\AA$)>0.15$\AA, on scales corresponding
to 200~km~s$^{-1} < \Delta v < 600$~km~s$^{-1}$, we measure a two-point
cross-correlation of $\xi = -1^{+1.75}_{-0}$ versus a corresponding single
sightline auto-correlation value of $\xi = 11.5\pm1.3$ \cite{SSB}.
This is most easily understood in terms of internal structure within absorbers
on velocity scales up to 600~km~s$^{-1}$, a novel result.

\vfill
\end{document}